\documentclass[amsmath,showkeys, amssymb,a4paper,11pt]{revtex4}
\usepackage{latexsym}
\usepackage[dvipdfm,colorlinks=true,linkcolor=cyan,anchorcolor=cyan,citecolor=cyan]{hyperref}
\usepackage[hmargin={2.54cm,2.54cm},vmargin={3.17cm,3.17cm}]{geometry}
\usepackage{float}
\usepackage{graphicx,amsfonts,amsmath,amssymb,amsthm,amscd}
\usepackage{epsfig}
\usepackage{url}
\usepackage{subfigure}
\usepackage{bm}
\usepackage{xr}
\usepackage{indentfirst}

\def \w{\omega_0}
\newcommand{\Eref}[1]{Eq.\,(\ref{#1})}
\newcommand{\Fref}[1]{Fig.\,\ref{#1}}

\graphicspath{{fig/}}

\date{\today}

\begin{document}

\title{
  \bf Theoretical model of transcription based on torsional mechanics of DNA template
}

\author{Xining Xu and Yunxin Zhang} \email[Email: ]{xyz@fudan.edu.cn}
\affiliation{Laboratory of Mathematics for Nonlinear Science, Shanghai Key Laboratory for Contemporary Applied Mathematics, Centre for Computational Systems Biology, School of Mathematical Sciences, Fudan University, Shanghai 200433, China.}

\begin{abstract}
Transcription is the first step of gene expression, in which a particular segment of DNA is copied to RNA by the enzyme RNA polymerase (RNAP). Despite many details of the complex interactions between DNA and RNA synthesis disclosed experimentally, much of physical behavior of transcription remains largely unknown. Understanding torsional mechanics of DNA and RNAP together with its nascent RNA and RNA-bound proteins in transcription maybe the first step towards deciphering the mechanism of gene expression. In this study, based on the balance between viscous drag on RNA synthesis and torque resulted from untranscribed supercoiled DNA template, a simple model is presented to describe mechanical properties of transcription. With this model, the rotation and supercoiling density of the untranscribed DNA template are discussed in detail. Two particular cases of transcription are considered, transcription with constant velocity and transcription with torque dependent velocity. Our results show that, during the initial stage of transcription, rotation originated from the transcribed part of DNA template is mainly released by the rotation of RNAP synthesis. During the intermediate stage, the rotation is usually released by both the supercoiling of the untranscribed part of DNA template and the rotation of RNAP synthesis, with proportion depending on the friction coefficient in environment and the length of nascent RNA. However, with the approaching to the upper limit of twisting of the untranscribed DNA template, the rotation resulted from transcription will then be mainly released by the rotation of RNAP synthesis.
\end{abstract}


\maketitle

\section{Introduction}
	Essential for all cell functions, transcription is the first step of gene expression \cite{Cooper2000}. It begins with the binding of RNA polymerase (RNAP) to a specific DNA sequence, which is usually named promoter \cite{Dehaseth1998,Cheetham1999RNApromoter}. As transcription proceeds, RNAP translocates along the template strand and uses base pairing complementarity with the DNA template to create an RNA copy \cite{Watson2013}.

	Due to the nature of DNA's right-handed double-helical structure \cite{Wilkins1953Molecular}, transcription requires a relative rotation of RNAP and its nascent RNA around DNA template. With elongation of RNA, however, it might become much easier to rotate a segment of the DNA being actively transcribed around its axis than to rotate the RNAC (short for RNA complex including RNA polymerase, its nascent RNA and any other proteins needed in transcription) \cite{Liu1987Supercoiling,Giaever1988Supercoiling}.
As described in the well-known twin-supercoiled-domain model \cite{Liu1987Supercoiling}, the advancing RNAC generates overwound (or positively supercoiling) for DNA downstream in transcription whereas DNA upstream becomes underwound (or negatively supercoiling), which has been validated in
both {\it in vivo} and {\it in vitro} studies as well \cite{Tsao1989Transcription,Krasilnikov1999Large,Harada2001Direct,Kouzine2004The,Forth2013Torque,Jie2016DNA}.

Recently, many efforts have been made to establish a full physical characterization of the transcription process. With combination of the simple free energy model of plectonemic supercoil \cite{Rybenkov1997} and the free energy of extended twisted DNA \cite{Moroz1997}, an analytical theory for the coexistence of extended and supercoiled DNA came to light, providing closed-form expressions for torque as a function of force and linking number \cite{Marko2007Torque}. Afterwards, \cite{Lavelle2014Pack} highlighted the significance of DNA supercoling as a major impact on gene expression. Then a framework describing the coupled RNA elongation, RNAP rotation and DNA supercoiling was introduced in  \cite{Sevier2017Mechanical}.

	In this study, inspired by the framework given in \cite{Sevier2017Mechanical}, a general model to describe the mechanical properties of transcription will be presented. With this model, the supercoiling density of DNA, rotation of RNAC, as well as the transcription velocity along DNA template can be obtained. Roughly speaking, for a long DNA template, transcription velocity and torque resulted from the supercoiling of DNA will reach their limit values as transcription going on. Meanwhile, with the motion of RNAP along DNA template, the total rotation shared by DNA will first increase rapidly, then keep almost constant, and finally decrease rapidly to zero.

   The organization of this study is as follows. The model and corresponding
theoretical analysis will be given in \autoref{Sec2}, and then results obtained by numerical calculations will be presented in \autoref{Sec3}. Finally, conclusions and remarks will be given in \autoref{Sec4}.

	\section{Theoretical model based on torsional mechanics}\label{Sec2}
	As stated in \cite{Sevier2017Mechanical}, the relative rotation of RNAC $\theta(x)$ and DNA sequence $\phi(x)$ are tied as
		\begin{equation} \label{Rotation}
		\omega_0 x = \phi(x)+\theta(x).
		\end{equation}
	Here, $x$ is the spatial distance of RNAP along gene template
away from the transcription start site (TSS), with $x>0$ for RNAP downstream from the TSS. The linking number of DNA $\omega_0 = 1.85   \text{ nm}^{-1}$ converts distance into angle of rotation, since the two strands of DNA in relaxed state wind around each other once approximately every 10.5 base pairs(bp), about 3.6 nm, forming a right-handed double helix \cite{Marko2007Torque}. The amount of $\phi(x)$ and $\theta(x)$ are related by the following balance
        \begin{equation} \label{eq2}
		\tau (\phi) = \Gamma(x,\dot{\theta}),
		\end{equation}
where $\tau (\phi)$ is the torque resulted from the supercoiling of DNA template, and $\Gamma(x,\dot{\theta})$ is the friction drag due to the viscous environment. Generally, $\Gamma(x,\dot{\theta})$ can be obtained by $\Gamma(x,\dot{\theta})=\gamma(x)\dot{\theta}$, with $\dot{\theta}$ the rotation speed and $\gamma(x)$ the rotational friction coefficient.

    It is well known that the rotational friction coefficient $\gamma$
depends on the shape and size of RNAC, see \cite{Howard2001}. With the forward motion of RNAP along DNA template, the length of nascent RNA will increase linearly with distance $x$. To describe this dependence, this study assumes that the friction coefficient $\gamma$ of RNAC is a function of $x$, and denoted as $\gamma(x)$. According to the usual linear approximation between friction drag $\Gamma$ and rotation speed, we get $\Gamma = \gamma(x)\dot{\theta}$. Particularly, in Ref. \cite{Sevier2017Mechanical}, $\gamma(x)=cx^{\alpha}$ is used.

In the following, we approximate the unknown function $\gamma(x)$ as $\eta(1+kx)$, with two parameters $\eta$ and $k$. Actually, $\eta(1+kx)$ can be regarded as the first two terms of the Taylor expansion of $\gamma(x)$. So, the friction drag $\Gamma$ can be approximated as $\Gamma = \eta(1+kx)\dot{\theta}$, in which the length-independent term $\eta \dot{\theta}$ can be regarded as the drag resulted from RNAP and other bounded proteins needed in transcription, and the linear term $\eta kx\dot{\theta}$ can be regarded as the drag resulted from the nascent RNA with length proportional to distance $x$.

By the chain rule of derivative, $\dot{\theta}=\partial_t\theta(x(t))=\theta^{\prime}\dot{x}=v\theta^{\prime}$. Where $v=\dot{x}$ is the elongation speed of RNA.
With identities in \Eref{Rotation} and \Eref{eq2}, we have
	
		\begin{equation}\label{main1}
		\tau(\phi) = \eta(1+kx)\dot{\theta} = \eta v (1+kx)(\omega_0-\frac{d \phi}{d x}) \,.
		\end{equation}

In \cite{Sevier2017Mechanical}, it is assumed that the twisting strain of DNA sequence at the point of transcription, which is caused by the forward motion of RNAP, spreads immediately throughout the specified DNA length. That is to say, the twisting is assumed to be shared uniformly by the untranscribed DNA sequence with length $L-x$, where $L$ denotes the total length of the coding sequence. This assumption is reasonable for slow transcription process, or transcription along a DNA template with large stiffness. To describe more general cases, this study takes the spread process of twisting strain into account and assumes that the DNA rotation (overwinding) $\phi$ is shared by the untranscribed DNA sequence with length $L-x$, but through an exponential distribution function with a parameter $\lambda$. Mathematically, the supercoiling density at the downstream position $s$ from the present transcription position $x$ can be expressed as $\sigma(s;x)= \sigma(0;x) \exp(-\lambda s)$. Considering the total rotation $\phi(x)$ of DNA sequence, $\sigma(s;x)$ should satisfy
		\begin{equation}\label{phi}
		\int_0^{L-x} \sigma(s;x) ds =\int_0^{L-x} \sigma(0;x) e^{-\lambda s} ds = \frac{\phi(x)}{\omega_0} \,,
		\end{equation}
which then gives the supercoiling density of DNA at the present transcription position $x$ as follows,
		\begin{equation} \label{sigma0}
		\sigma(0;x) = \frac{\lambda}{\omega_0}\phi(x)(1-e^{-\lambda(L-x)})^{-1} \,.
		\end{equation}

Generally, parameter $\lambda$ may depend on the dwell time of RNAP at each base pair of DNA template, or equivalently depend on the transcription velocity $v(x)$. For convenience of theoretical analysis, this study only discusses the simple cases in which parameter $\lambda$ is constant, this is reasonable if transcription is almost in steady state.

For limit case $\lambda \rightarrow 0$, $\sigma(s;x)$ reduces to an uniform distribution, $\sigma(s;x)\equiv\sigma(0;x)= \phi(x)/\w(L-x)$, which is just the one discussed in \cite{Sevier2017Mechanical}. For simplicity, we denote $\sigma(0;x)$ as $\sigma(x)$ in the following. \Eref{sigma0} can be reformulated as
		\begin{equation} \label{sigma01}
        \phi(x) = \omega_0 \sigma(x) (1-e^{-\lambda(L-x)})/\lambda \,.
		\end{equation}
Substitution of \Eref{sigma01} into \Eref{main1} gives
		\begin{equation} \label{main2}
		\lambda \tau(\sigma) = \w \eta v(1+kx)\left (\lambda-[\sigma(x)p(x)]^{\prime} \,\right) \, ,
		\end{equation}
with $p(x):=1-e^{-\lambda(L-x)}$.

    According to Ref. \cite{Marko2007Torque}, the supercoiling density
$\sigma(x)$ dependent torque $\tau$ resulted from the overwinding of DNA template, as appeared in \Eref{main2}, can be approximated as follows,
		\begin{equation}\label{tau}
			\tau(\sigma) =
			\left \{
			\begin{aligned}
			S\sigma , \quad & \sigma < \sigma_s\, , \\
			\tau_0, \quad &\sigma_s< \sigma < \sigma_p\, , \\
			P\sigma, \quad & \sigma_p < \sigma\, .
			\end{aligned}
			\right.
		\end{equation}
Where parameters $S$, $\tau_0$, $P$ and supercoiling transition values $\sigma_s,\, \sigma_p$ are determined by mechanical properties of DNA. Phenomenically, $\sigma_s$ and $\sigma_p$ are supercoiling densities when DNA exceeds stretched state and reaches plectonemic, respectively.

	The relation $\tau(\sigma)$ shown in \Eref{tau} formulates \Eref{main2} as two types of equations. For constant torque $\tau\equiv\tau_0$, {\it i.e.}, when supercoiling density $\sigma$ lies between $\sigma_s$ and $\sigma_p$, \Eref{main2} reduces to
		\begin{equation}\label{eqt1}
		\lambda \tau_0 = \w \eta v(1+kx)(\lambda-[\sigma(x)p(x)]^{\prime}) .
		\end{equation}
	While for the other two linear relation cases, {\it i.e.}, when $\sigma < \sigma_s$ or $\sigma > \sigma_p$, \Eref{main2} reduces to
		\begin{equation}\label{eqt2}
		\lambda W \sigma = \w \eta v(1+kx)(\lambda-\sigma^{\prime}(x)p(x)-\sigma(x)p^{\prime}(x))\,,
		\end{equation}
	with $W = S$ or $W = P$ respectively, see \Eref{tau}.

    For cases with constant transcription velocity, {\it i.e.}, $v \equiv v_0$,
the first type equation (\Eref{eqt1}) can be solved by direct integration, while for the other type of equation (\Eref{eqt2}) it can be solved by multiplying an integrating factor $\mu_{W}(x)=\exp \left(\lambda \widetilde{W}\int[(1+kx)p(x)]^{-1}dx \right)$, with $\widetilde{W} := W/(\w \eta v_0)$. These lead to the full solution for supercoiling density as follows,
	\begin{equation}
	\label{ode0}
	\sigma(x) = \left\{
	\begin{aligned}
	&\left.\left(\lambda\int_0^x \mu_{S}(y) \, dy\right)\right/p(x)\mu_{S}(x) \, ,&\quad \sigma < \sigma_s,\, \\
	&\left.\left(\lambda x - \frac{\lambda \widetilde{\tau}_0}{k}\ln(1+kx) + C_1\right)\right/p(x), &\quad \sigma_s < \sigma < \sigma_p \\
	& \left.\left(\lambda \int_0^x \mu_{P}(y) \, dy + C_2 \right)\right/p(x)\mu_{P}(x)\, ,&\quad \sigma > \sigma_p \, ,
	\end{aligned}
	\right. \, ,
	\end{equation}
where $\widetilde{\tau}_0 := \tau_0/(\w \eta v_0)$, and constants $C_1, C_2$ can be determined by matching boundary conditions to keep $\sigma(x)$ continuous at $\sigma=\sigma_s$ and $\sigma=\sigma_p$. Finally, the rotation of DNA $\phi(x)$ and torque $\tau(x)$ can be obtained by Eqs. (\ref{sigma01},\,\ref{tau}) and Eq. (\ref{ode0}), and by \Eref{Rotation} the rotation of RNAC $\theta(x)=\omega_0x-\phi(x)$ can also be obtained.

Experiments show that positive supercoiling built up by transcription is a impediment that may slow down the motion of  RNAC \cite{Chong2014Mechanism}. So, generally a {\it stalling torque} $\tau_c$ of RNAP, under which the transcription velocity $v$ is vanished, should be introduced. To describe these more general cases, similar as the discuss in motor proteins \cite{Howard2001,Lipowsky2008,Kunwar2010Robust}, the following velocity-torque relation of RNAP is assumed,
	\begin{equation} \label{eqv}
	v = v_0\left[1 - \left(\tau(\sigma)/\tau_c\right)^n\,\right] \,,
	\end{equation} 		
with $v_0$ the torque-free velocity of RNAP. Take \Eref{eqv} into \Eref{eqt1} and \Eref{eqt2}, we obtain {\small
	\begin{align}
		\label{ode11}
	\lambda \widetilde{\tau}_0  \left[1-\left(\tau_0/\tau_c\right)^n \, \right]^{-1} & =(1+kx)(\lambda-[\sigma(x)p(x)]^{\prime}) ,
	\quad & \sigma_s < \sigma < \sigma_p,  \\
		\label{ode12}
	\lambda \widetilde{W} \sigma(x)\left[ 1-\left(W\sigma/\tau_c\right)^n\,\right]^{-1}&=
	(1+kx) (\lambda - \sigma^{\prime}(x)p(x)-\sigma(x)p^{\prime}(x)),\ & \sigma<\sigma_s\ \textrm{or}\ \sigma >\sigma_p\, ,
	\end{align}
}
with $\widetilde{\tau}_0, \widetilde{W}$ defined the same as in \Eref{ode0}.

\Eref{ode11} is different from \Eref{eqt1} only in the left-hand side through an additional constant $[1-(\tau_0 /\tau_c)^n]^{-1}$, so its solution can be obtained explicitly. But for \Eref{ode12}, it is much difficult to obtain its solution for general exponent $n$. We will solve them numerically with the similar boundary conditions as for Eqs. (\ref{eqt1},\,\ref{eqt2}) in the following section.

	\section{Rotation and torque of DNA template during transcription }\label{Sec3}
	\begin{figure}[!htbp]
		\centering
		\includegraphics[width=0.8\textwidth]{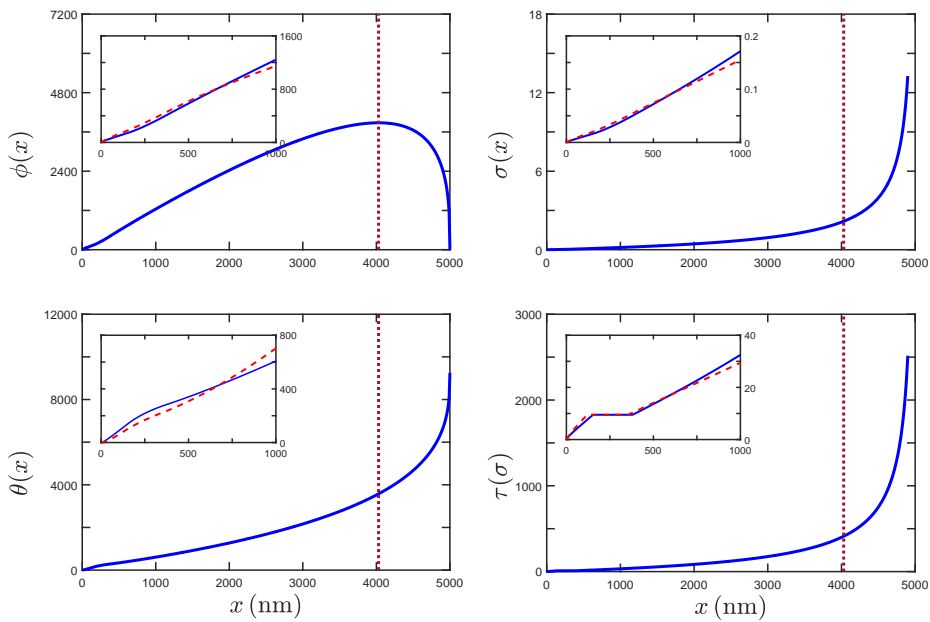}
		\caption{Rotation $\phi$, torque $\tau$, supercoiling density $\sigma$ of DNA template, and rotation $\theta$ of RNAP, as a function of length $x$ of the nascent RNA. Parameter values used in calculations are $L=5000$ nm, $\eta v_0=1$, $k=0.055$, $\lambda = 10^{-5}$, and $S = 582.0$, $\tau_0=9.5$, $P=189.6$ pN nm \cite{Sevier2017Mechanical}. Supercoiling density $\sigma(x)$ is calculated from \Eref{ode0}, rotation $\phi(x)$ is calculated from \Eref{sigma01}, rotation $\theta(x)$ is calculated by $\theta(x)=\omega_0 x-\phi(x)$ (see \Eref{Rotation}), and torque $\tau$ is calculated from \Eref{tau}. The vertical dotted line corresponds to the position at which rotation $\phi(x)$ reaches its maximum. The dashed lines in inserted zoomed figures are calculated from the model employed in \cite{Sevier2017Mechanical} for comparison, where friction drag on RNAC is calculated by $\Gamma = \eta x^{1/2}\dot{\theta}$, and rotation $\phi(x)$ of DNA template is always assumed to be shared equally by the untranscribed part with length $L-x$.
		}\label{Result0}
	\end{figure}

    In this section, based on theoretical models given in Sec. \ref{Sec2},
rotation of RNAC $\theta$ and DNA template $\phi$, as well as the torque $\tau$ and supercoiling density $\sigma$ of DNA template will be discussed in detail. Firstly, for the simple cases with constant transcription velocity $v\equiv v_0$, the results are displayed in \Fref{Result0}, which are calculated from \Eref{eqt1} and \Eref{eqt2}, or \Eref{ode0} equivalently. Results show that, with the forward motion of RNAP, the rotation of DNA $\phi(x)$ will increase first and then decrease to zero rapidly, though the supercoiling density $\sigma(x)$ and torque $\tau(x)$ always increase. For short nascent RNA ($x\lesssim750$ nm), the slope of $\theta(x)$ decreases while that of $\phi(x)$ increases with $x$, which means that the proportion of total rotation $\omega_0x$ shared by DNA/RNAP increases/decreases with $x$. One can image that, initially most of the rotation will be shared by RNAP since there is no nascent RNA and therefore the friction drag $\Gamma$ on RNAP is very small. With the elongation of RNA, the friction drag $\Gamma$ becomes larger compared with the strain in DNA sequence, then more and more rotation will be shared by DNA sequence. However, contrary to previous cases, for long nascent RNA ($x\gtrsim750$ nm), the slope of $\theta$ increases while that of $\phi(x)$ decreases with $x$, which means that the proportion of total rotation $\omega_0x$ shared by DNA/RNAP decreases/increases with $x$. The reason is that, with increase of $x$, the friction drag $\Gamma$ on RNAC increases only linearly while the strain on DNA sequence increases nonlinearly (with order larger than 1). Therefore, more and more rotation have to be shared by the RNAC. Which has been observed previously in \cite{Liu1987Supercoiling}. After a critical value $x^*$ (corresponding to the dotted vertical line in \Fref{Result0}), the slope of $\phi(x)$ will change from positive to negative and with the slope of $\theta(x)$ increasing rapidly. This implies that, for $x>x^*$, all of the newly produced rotation $\omega_0(x-x^*)$ will be shared by RNAC. Meanwhile, the rotation previously stored in DNA's double helix will also be partly shared by RNAC, since the downstream DNA's double helix almost reaches it twisting limit. But the supercoiling density $\sigma(x)$ and torque $\tau$ will never decrease. Finally, all rotation of DNA template $\omega_0L$ will be released by the rotation of RNAC, and with $\phi(L)=0$. Due to their linear relationship (see \Eref{tau}), torque $\tau(x)$ and supercoiling density $\sigma$ have similar behavior, except that $\tau(x)$ is a piecewise function with three pieces. Finally, dashed lines plotted in the inserted zoomed subfigures of \Fref{Result0} are calculated by the model used in \cite{Sevier2017Mechanical} with friction drag given by $\Gamma = \eta x^{1/2}\dot{\theta}$, with the assumption that rotation in the untranscribed DNA template is distributed uniformly, or equivalently the untranscribed template with length $L-x$ is always in steady state. It shows that the model used in \cite{Sevier2017Mechanical} can be regarded as the limited case of the one presented in this study, with parameter $\lambda$ tending to zero.
	
	\begin{figure}[!htbp]
		\centering
		\includegraphics[width=0.8\textwidth]{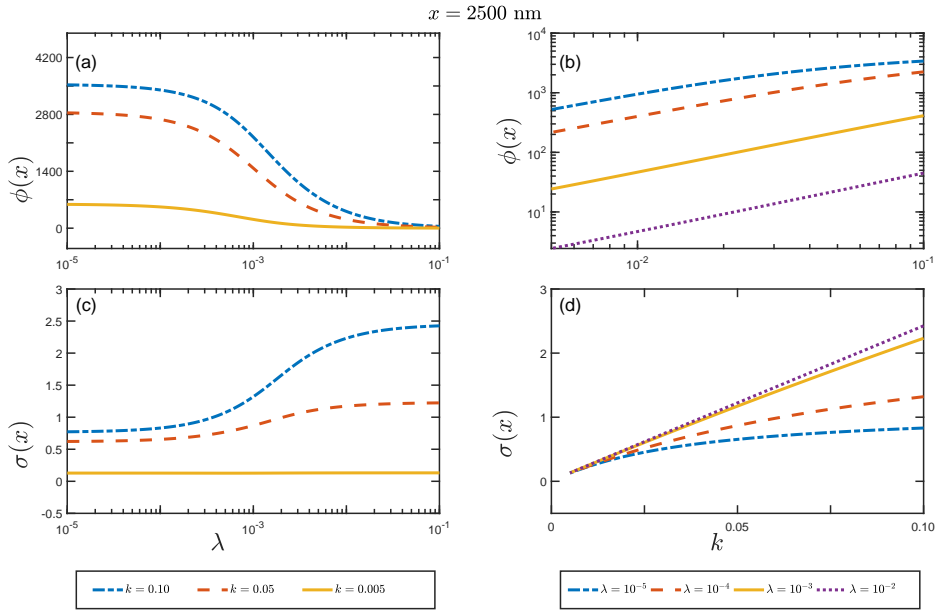}
		\caption{DNA rotation $\phi$ and supercoiling density $\sigma$ when RNAP reaches a particular position at $x = 2500$ nm away from the transcription start site, {\it i.e.}, the midpoint of a DNA sequence of length $L=5000$ nm. Here, $S$, $\tau_0$ and $P$ are kept the same as those used in \Fref{Result0}. (a) and (c) show how $\phi$ and $\sigma$ change with parameter $\lambda$ for $k = 0.1$(dash-dot lines), $0.05$(dashed lines) or $0.005$(solid lines). While (b) and (d) illustrate the behavior of $\phi$ and $\sigma$ as functions of parameter $k$ for four typical values of $\lambda$ (dash-dot lines for $\lambda = 10^{-5}$, dashed lines for $\lambda = 10^{-4}$, solid lines for $\lambda = 10^{-2}$, and dotted lines for $\lambda = 10^{-1}$ ).}
		\label{DP0}
	\end{figure}

The influences of parameters $k$ and $\lambda$ (see \Eref{main1} and \Eref{phi}) on the mechanical properties of DNA template are illustrated with \Fref{DP0}. Obviously, the drag on RNAC $\Gamma(x,\dot{\theta})=\eta v(1+kx)\dot{\theta}$ increases with $k$. For large $k$, more rotation will be shared by the untranscribed DNA sequence, so rotation $\phi(x)$ and supercoiling $\sigma(x)$ increases with parameter $k$. Moreover, our results surprisingly show that $\phi\approx\alpha k^{\beta}$, with $\alpha$ decreasing with $\lambda$ while $\beta$ almost independent of $\lambda$, see \Fref{DP0}(b). For large parameter $\lambda$, the twisting strain of DNA helix will concentrate more at exactly the current transcriptional position $x$. Therefore, it will become more difficult to rotate the DNA helix, and consequently, more rotation will be released by RNAC, which means rotation $\phi(x)$ of DNA decreases with parameter $\lambda$, see \Fref{DP0}(a). However, the supercoiling density $\sigma(x)$ at current transcription position increases with parameter $\lambda$, since with large $\lambda$ more rotation will be concentrated at the transcription point (see \Fref{DP0}(c)), though the total rotation $\phi(x)$ shared by all the untranscribed DNA helix decreases with $\lambda$. As mentioned before, for $\lambda\to 0$ the model presented in this study reduces to the simple case as employed in \cite{Sevier2017Mechanical}. Meanwhile, plots in \Fref{DP0}(a,\, c) show that both $\phi(x)$ and $\sigma(x)$ become insensitive to parameter $\lambda$ for large or small values of $\lambda$, {\it i.e.}, there are limit values for $\lambda\to0$ or $\lambda\to\infty$. The limit value of $\phi(x)$ for large $\lambda$ is zero, which means that, when $\lambda$ is large enough, almost all rotation resulted from transcription will be released by RNAC. However, for $\lambda\to\infty$, the limit value of supercoiling density $\sigma$ is nonzero, which increases with parameter $k$. This is because that, with the increase of $k$, the drag $\Gamma$ on RNAC will also increase. Therefore, the twist strain needed to balance the drag should increase, see \Eref{eq2}. Similarly, the limit values of $\phi(x)$ and $\sigma(x)$ for $\lambda\to 0$ also increase with parameter $k$. Finally, the plots in \Fref{DP0}(d) show that, for large parameter $\lambda$, supercoiling density $\sigma(x)$ increases almost linearly with $k$. The reason is that, for large $\lambda$ almost all the twisting strain needed to balance the drag $\Gamma$ on RNAC concentrates on the current transcription position $x$, while for given value of $x$ the drag $\Gamma$ increases linearly with parameter $k$. It is noted that, due to the linear relationship between $\tau$ and $\sigma$ (see \Eref{tau}), the dependence of torque $\tau$ on parameters $\lambda$ and $k$ is similar as that of $\sigma$, and therefore is not shown in \Fref{DP0}. Finally, for more detailed calculations with regard to the constant transcription velocity cases, see Figs. S1-S4, and Tab. SI in supplementary materials.

	\begin{figure}[!htbp]
		\centering
		\includegraphics[width=0.8\textwidth]{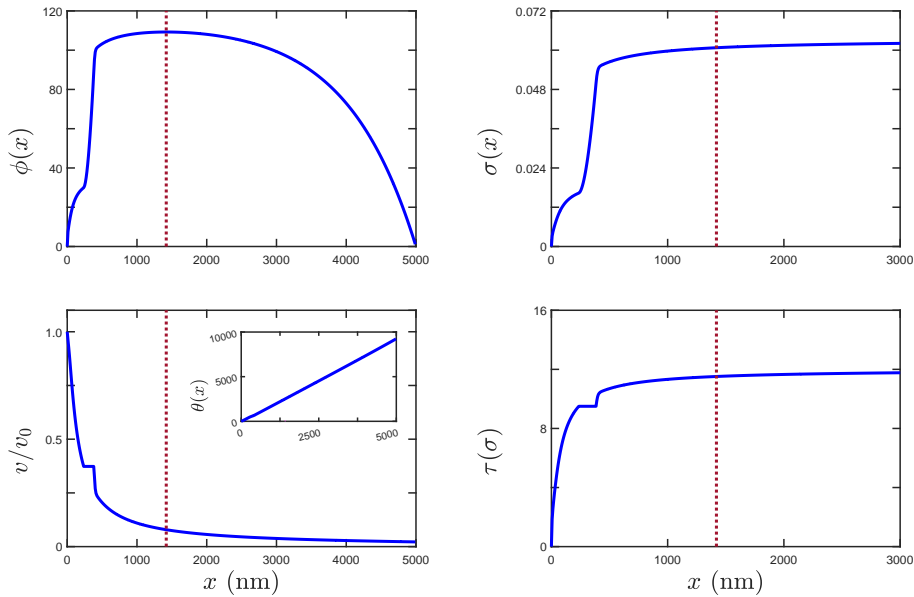}
		\caption{Rotation $\phi(x)$, supercoiling density $\sigma(x)$, torque $\tau(x)$ of DNA template, and the transcription velocity $v$ for general cases in which velocity $v$ is torque dependent, see Eqs. (\ref{eqv},\, \ref{ode11},\, \ref{ode12}). Parameter values used in calculations are $L=5000$ nm, $k = 0.055$, $\lambda = 10^{-3}$, $n=2$, and $\eta v_0$, $S$, $\tau_0$, $P$ are kept the same as those in \Fref{Result0}. The same as  \cite{Ma2013Transcription}, {\it stalling torque} of the RNAP is assumed to be $\tau_c = 12$ pN nm. Similar to \Fref{Result0}, dotted lines show the position $x$ at which DNA rotation $\phi$ reaches its maximum.}
		\label{Result1}
	\end{figure}

As mentioned in \cite{Ma2013Transcription}, the assumption that RNAP always moves along DNA template with a constant velocity is not generally true, especially for real transcription {\it in vivo}. For example, supercoiling density $\sigma$, and consequently the torque $\tau$, will become pretty large as RNAP approaches the end of template, especially for RNAP working in a viscous environment. With a particular form of velocity-torque relation given in \Eref{eqv}, results of rotation $\phi(x)$, supercoiling density $\sigma(x)$, and torque $\tau(x)$ of DNA template are plotted in \Fref{Result1}. Similar as the constant velocity cases, rotation $\phi(x)$ increases first and then decreases to zero with the completion of transcription, and both $\sigma(x)$ and $\tau(x)$ increase monotonically with the length $x$ of nascent RNA. But an obvious difference is that rotation $\phi(x)$ in \Fref{Result1} increases more rapidly than that in \Fref{Result0}, while decreases more slowly than that in \Fref{Result1}. With the increase of length $x$ of nascent RNA, twist strain induced by the overwinding of DNA helix will increase. This implies that the transcription velocity $v$ decreases with $x$, see the left-bottom figure in \Fref{Result1}. Meanwhile, \Fref{Result1} shows that, during the initial period of transcript, the torque $\tau(x)$ increase also very rapidly. Thus, the drag on RNAC $\Gamma=\eta(1+kx)v\theta'$ increases very rapidly with $x$ during the initial period. In some sense, \Fref{Result1} shows that, with torque dependent velocity, transcription (especially the physical state of DNA template) will reach its steady state more rapidly. In the initial period of transcription, most of the rotation will be shared by DNA helix until it approximates its upper limit of capacity, see the plots in \Fref{Result1} for torque $\tau$ and supercoiling density $\sigma$.

	\begin{figure}
		\centering
		\includegraphics[width=0.8\textwidth]{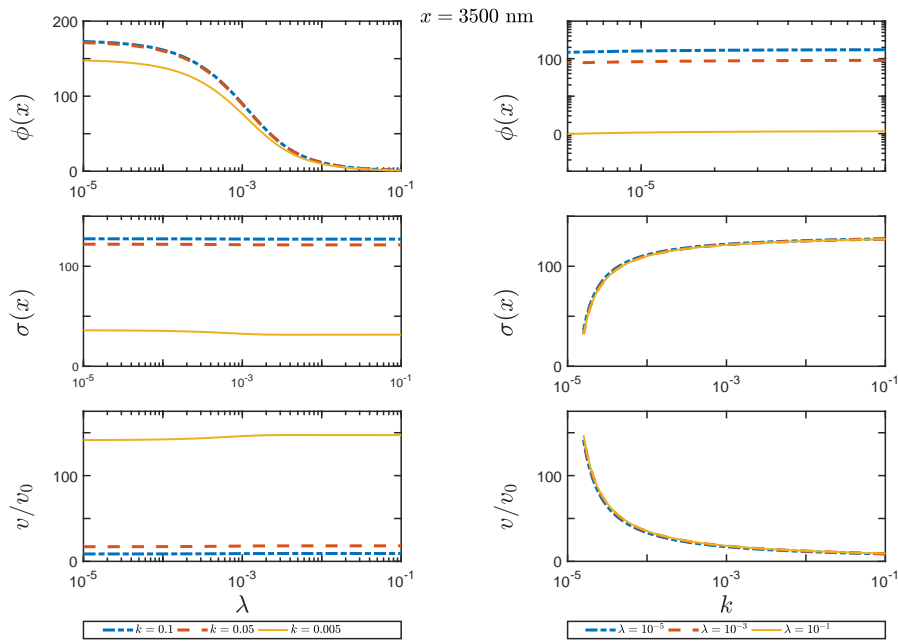}
		\caption{Rotation $\phi$, supercoiling density $\sigma$ of DNA helix, and velocity $v$ of RNAP, at a typical position which is $x=3500$ nm away from the transcription start site. The parameter values used in calculations are $L = 5000$ nm, $\tau_c=12$ pN nm, $\eta v_0 = 1$, $S = 582.0$ pN nm, $\tau_0 = 9.5$ pN nm and $P = 189.6$ pN nm. The $\lambda$ dependent results are shown on the left with $k=0.1,\, 0.05,\, 0.005$, respectively. The $k$ dependent results are shown on the right with $\lambda = 10^{-5},\, 10^{-3},\, 10^{-1}$ respectively.}
		\label{DP1at3500}
	\end{figure}

Similar to \Fref{DP0}, dependent properties of parameters $\lambda$ and $k$ obtained by the general model with torque dependent velocity of RNAP are plotted in \Fref{DP1at3500}. The dependence of DNA rotation $\phi$ on parameter $\lambda$ is similar to that in \Fref{DP0}, the constant transcription velocity cases, which changes from one limit constant to zero as the increase of $\lambda$. But the supercoiling density $\sigma$ is almost independent of $\lambda$, especially for large $k$ cases. Consequently, the velocity $v$ is almost independent of $\lambda$, see \Eref{eqv}. These imply that, with the increase of $\lambda$, the supercoiling density $\sigma$, or equivalently the torque $\tau$ needed to balance the drag $\Gamma$ on RNAC, is almost constant at current transcription site, though the total rotation $\phi$ of the untranscribed DNA helix decreases. Or in other words, although the total rotation $\phi$ of DNA decreases with $\lambda$, the supercoiling density $\sigma$ at the current transcription position is almost independent of $\lambda$, since large parameter $\lambda$ means more proportion of rotation $\phi$ will be concentrated at the current transcription position. In short, two factors that might influence the value of $\sigma$ balance each other. On the other hand, the three right subfigures in \Fref{DP1at3500} show that rotation $\phi$ is almost independent of parameter $k$, while supercoiling density $\sigma$ increases slightly with $k$. With large value of $k$, the drag $\Gamma=\eta(1+kx)\dot{\theta}$ on RNAC will be large, therefore the torque $\tau$ induced by the overwinding of DNA helix, which is needed to balance $\Gamma$, should be large. One possible reason that $\phi$ is almost independent of $k$ and $\sigma$ is almost independent of $\lambda$ is that, at position $x=3500$ nm, the rotation $\phi$ and supercoiling density $\sigma$ have almost reached their upper limits, and therefore are not sensitive to the change of drag $\Gamma$ and parameter $\lambda$, respectively. For calculation results with $x=100$ nm, see Fig. S5 in supplementary materials.

		\section{Conclusions and Remarks}\label{Sec4}
In this study, theoretical model to describe the mechanical properties of transcription process is presented. Roughly speaking, our model is mainly based on the balance between the friction drag on RNAC, which is caused by the viscous environment in cells, and the torque induced by the overwinding of DNA helix caused by transcription. As usual, the drag is calculated by the product of friction coefficient and rotation velocity, but with the friction coefficient depending on the length of nascent RNA, as discussed in \cite{Harada2001Direct}. Meanwhile, the torque is calculated by the supercoiling density of DNA helix. The basic idea of our model is similar to the one used in \cite{Sevier2017Mechanical}, while the methods to calculate the friction drag on RANC and the supercoiling density of DNA helix are generalized. Two types of transcription are discussed in this study, one with constant transcription velocity and the other with torque dependent velocity. Our results show that, during the initial period of transcription, the rotation coming from the transcribed part of DNA template is mainly shared by the untranscribed part of DNA template. With the increase of supercoiling density and twisting strain induced by overwinding of the untranscribed part of DNA template, the rotation resulted from transcription will then be mainly released by the rotation of RNAC. But anyway, all the rotation of DNA template will finally be totally released by the rotation of RNAC. Although the model presented in this study still looks too simple to get detailed properties of transcription quantitively, it is reasonable enough to understand the basic principle of transcription, especially the mechanism of how to deal with the natural rotation stored in DNA template during the transcription process. It may need to note that the model presented in this study can also be generalized to describe the transcription process with multiple (cooperated) RNAPs aligning along one single DNA template, which may be helpful to the understanding of transcriptional bursting in gene expression \cite{Chong2014Mechanism,Nicolas2017What}.


\end{document}